\documentclass[reprint,nobibnotes,superscriptaddress]{revtex4-1}

\usepackage{graphicx}
\usepackage{amsmath}
\usepackage{amsfonts}
\usepackage{amssymb}

\usepackage{natbib}
\usepackage[normalem]{ulem}
\usepackage[table,xcdraw]{xcolor}
\usepackage{epsfig}
\usepackage{color}
\usepackage{bm}
\usepackage{dashrule}
\usepackage{hyperref}
\hypersetup{
	colorlinks=true,
	linkcolor=blue,
	filecolor=magenta,      
	urlcolor=red,
	citecolor=blue,
}
\usepackage{todonotes}
\newcommand{\minus}{\scalebox{0.75}[1.0]{$-$}}

\newcommand{\pdag}{{\phantom\dagger}}

\begin{document}
	
\title{Softening of $dd$ excitation in the resonant inelastic x-ray scattering spectra as a signature of Hund's coupling in nickelates}

\author{Umesh Kumar}
\affiliation{Department of Physics and Astronomy, Rutgers University, Piscataway, NJ 08854, USA}
\author{Corey Melnick}
\affiliation{Condensed Matter Physics and Materials Science Department, Brookhaven National Laboratory, Upton, NY 11973, USA}
\author{Gabriel Kotliar}
\affiliation{Department of Physics and Astronomy, Rutgers University, Piscataway, NJ 08854, USA}
\affiliation{Condensed Matter Physics and Materials Science Department, Brookhaven National Laboratory, Upton, NY 11973, USA}

\begin{abstract}
We investigate the effects of Hund's coupling on the resonant X-ray absorption spectra of the recently discovered family of layered nickelate superconductors. We contrast two scenarios depending on the relative strength of the ratio of the effective Hund's coupling ($J_H$) to the crystal fields  ($\Delta$) in these systems. We carry out the cluster and DFT+DMFT  simulations of the RIXS signal at the   Ni $L$-edge for different values of Hund's coupling. We find the latter dominates for the parent compound while the former becomes important for sufficiently large doping. Our results are consistent with the observations of a softening of a RIXS peak as a function of doping by  Rossi {\sl et al.}~\cite{PhysRevB.104.L220505}, only when the Hund coupling is sizeable.   To interpret the results, we separate the theoretical RIXS signal into spin conserving and non-spin conserving 
channels and conclude that the infinite layer nickelates are in a regime where $\Delta$ and $J$ compete effectively and suggest further experimental tests of the theory. 
\end{abstract}	
\maketitle	

{\sl Introduction:} A long-standing goal in condensed matter physics has been to discover a superconducting infinite-layer (IL) nickelate, as these compounds are analogous to the high-temperature two-dimensional (2D) superconducting cuprates ~\cite{PhysRevB.59.7901}, in which superconductivity was discovered in 1986~\cite{Bednorz1986}.
After a long search, superconductivity was discovered in the doped infinite layer nickelate, Nd$_{0.8}$Sr$_{0.2}$NiO$_2$~\cite{Li2019}, with the differences in superconducting domes with the 2D cuprates~\cite{PhysRevLett.125.027001, doi:10.1146/annurev-conmatphys-070909-104117}.
Recently, superconductivity was also reported in La$_3$Ni$_2$O$_7$, suggesting it as a generic feature of nickelates~\cite{Sun2023}. This is a major breakthrough, as the electronic structure of the nickelates is notably different from the cuprates. Most notably, the nickel in a nickelate is understood to exist in electron configurations from $d^7$ to $d^9$, unlike the copper in a cuprate superconductor, which is understood to exist only in the $d^9$ configuration~\cite{10.3389/fphy.2021.808683, PhysRevLett.112.106404}.
Nickelates are also known to host metal-insulator transitions where the oxidation state of Ni plays a central role~\cite{PhysRevLett.112.106404}. 

Superconductivity in doped 2D cuprates is understood to be driven by resonating valence bond states~\cite{PWAnderson1987} and has $d$-wave pairing~\cite{PhysRevB.37.3664, PhysRevB.38.5142}. This is usually explored using the Emery model~\cite{PhysRevLett.58.2794}, which allows for Zhang-Rice singlets formed with $d_{x^2-y^2}$ orbitals~\cite{PhysRevB.37.3759} and the oxygens.
The parent compound for IL nickelate is weakly insulating in contrast to the Mott insulator realized in cuprates. The possibility of mixed oxidation states in nickelates makes the problem challenging, as multiple $d$ orbitals can play a significant role in the origin of the superconductivity. In particular, $3z^2$ have been hypothesized to play a role in the ground state~\cite{PhysRevLett.126.127401, PhysRevX.10.041002}. The model for nickelates ranges from single-band to the multiorbitals of the Ni atom, and even the role of the rare-earth metal has been considered
~\cite{PhysRevResearch.1.032046,PhysRevB.101.041104,PhysRevB.101.075107, PhysRevResearch.2.023112, PhysRevB.102.220501,PhysRevX.10.041047,PhysRevX.10.041002, PhysRevB.101.060504, Gu2020,PhysRevX.11.011050, PhysRevLett.124.207004, Kitatani2020}.

The Fermi surfaces of the rare earth metal have also been postulated to play a role in the IL nickelate NdNiO$_2$~\cite{Hepting2020, PhysRevX.11.041009, Wang2020}, which complicates the understanding of the active degrees of freedom for superconductivity.  The relevance of Hund's coupling has also been explored~\cite{Wang2020, Jin_2022, Chang2020, Kang2023}. More recently, it has been proposed that since these compounds are created by reduction, the hydrogen atoms in these materials can play a significant role, further complicating the description~\cite{Ding2023}. Indeed, the origin of superconductivity is a controversial topic, and the proposals are at a very nascent stage. 

Recently, resonant inelastic x-ray scattering (RIXS)~\cite{RevModPhys.83.705, WANG2019151, UKumar2022, Rossi2022, PhysRevB.104.L220505, YShen2022, YShen2023} has also been extensively used to explore the relevant degrees of freedom in nickelates. It offers an advantage as it can amplify the response for smaller cross-sections, such as small crystals and thin films, due to large light-matter interaction and inherent resonance conditions in the technique. The spectroscopic response allows one to constrain the possible theory for superconductivity in nickelates. For example, In a  recent study~\cite{YShen2022} on low valence nickelates, La$_4$Ni$_3$O$_8$ and La$_{2-x}$Sr$_{x}$CuO$_{4}$ revealed that the nickelates are of mixed charge-transfer–Mott-Hubbard character with the Coulomb repulsion $U$, in contrast to the charge-transfer nature of the cuprates in Zaanen-Sawatzky-Allen scheme. 
The distinction between $dd$ excitations and charge transfer (CT) excitations helps to reveal the differences between the two. Recently, a softening of $dd$ excitations in IL nickelate on doping was reported in Ni $L$-edge RIXS spectra~\cite{PhysRevB.104.L220505}. We revisit the observation to investigate the relevance of Hund's coupling to the observation.

In this letter, we explore the $dd$ excitations in undoped and doped nickelates. In the undoped case, $dd$  excitation, the transition between singly occupied $d_{x^2-y^2}$ and other $d$ orbitals is independent of $J_H$. In the doped case, there are two possible scenarios: i) $d_{x^2-y^2}$ is double occupied or ii) holes occupy $d_{x^2-y^2}$ and $d_{3z^2}$ depending on the relative strength of the splitting between $d$ orbitals ($\Delta$) and $J_H$, which leads to either softening or hardening of $dd$ excitations. Studying RIXS spectra using a small cluster and DFT+DMFT, we show that $J_H$ plays a central role in softening $dd$ excitations despite $\Delta$ dominating the ground state and report results consistent with the experimental softening of $dd$-excitations on doping~\cite{PhysRevB.104.L220505}.

\begin{figure}
\vspace{-0.0cm}
\centering
\includegraphics[width=\linewidth]{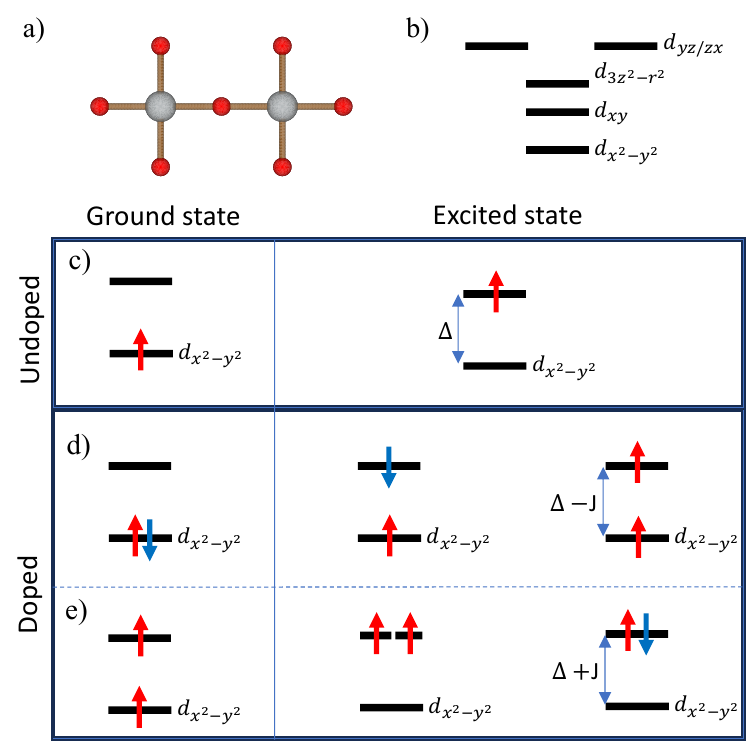}
\vspace{-0.25cm}
\caption{ Schematics of the cluster and the ground state occupancies in the cluster. a) shows the Ni$_2$O$_7$ cluster and b) shows its Ni $3d$ energy levels. c)-e) shows excitations for single ion configurations. c) shows schematics for the ground and $dd$ transition for the undoped model. d) and e) show the schematics for ground and the $dd$ transitions for the doped model with two ground state configurations; i) doubly occupied $d_{x^2-y^2}$ orbital for large $\Delta$, and ii) one hole each in $d_{x^2-y^2}$ and $d_{3z^2}$ orbitals for large $J_H$, respectively. In the spin-flip ($\Delta S= 1$) channel, d) and e) will soften and harden $dd$ excitations,  respectively. \label{fig:Schematicsdd}} 
\vspace{-0.5cm}
\end{figure}

{\sl Model:---} We consider the Ni$_2$O$_7$ cluster shown in Fig.~\ref{fig:Schematicsdd}(a) with all the Ni $3d$ and O $2p $ orbitals, unless otherwise stated. 
 The full Hamiltonian is given by $\mathcal{H}  =  \mathcal{H}_O+ \mathcal{H}_C^d+\mathcal{H}_C^p + \mathcal{H}_{pd}$~\cite{YShen2022}. The non-interacting part is  
\begin{equation*}
\begin{split}
&  \mathcal{H}_{0}  = \sum_{i\alpha}\epsilon_{\alpha} n_{i\alpha}^d + \sum_{i\mu}\epsilon_{\mu} n_{i\mu}^p+  \sum_{\langle i\alpha,j\mu\rangle} t_{i\alpha,j\mu} d_{i\alpha\sigma}^\dagger p_{j\mu\sigma}^{\phantom{\dagger}}  \\
&\qquad  + \sum_{\langle\langle i\mu,j\mu'\rangle\rangle, \sigma} t_{i\mu, j\mu'} p_{i\mu\sigma}^\dagger p_{j\mu'\sigma}^{\phantom{\dagger}} +\text{h.c.}
\end{split}
\end{equation*}
 Here $\alpha\in \{d_{x^2-y^2},d_{3z^2},d_{xy},d_{yz},d_{zx}\}$ Ni 3$d$ orbitals and  $\mu\in\{p_x, p_y, p_z\}$ O 2$p$ orbitals.
 $\epsilon_{\alpha/\mu}$ is the onsite energy at site \textit{i}. $t_{i\alpha,j\beta}$ is hopping between orbital  $\alpha$ of site \textit{i}  to $\beta$ of \textit{j}.  $ \mathcal{H}_{pd}  =  V_{pd}\sum_{ij}\ n_{i}^d n_{j}^p$.
 $V_{dp}$ is nearest neighbor coulomb interaction. The Coulomb interaction on O atom is given by $H_C^p  = U_p\sum_{i, \mu} n_{i\mu\uparrow}^p n_{i\mu\downarrow}^p $. $\mathcal{H}_C^d $ is the Coulomb repulsion on the $3d$ of Ni are accounted for by the multiorbital Hubbard-Kanamori model~\cite{Dagotto20011,PhysRevB.72.214431, Kumar2022} give by 
\begin{equation*} \label{eq:Coul}
\begin{split}
&\mathcal{H}_C^d =  U_d\sum_{ i,\alpha} n_{i\alpha\uparrow}^d n_{i\alpha\downarrow}^d +U_d^{\prime}\sum_{\substack{i, \sigma,\sigma', \\  \alpha <\beta }} n_{i\alpha \sigma}n_{i \beta \sigma'} \\
& + J_H \sum_{\substack{i, \sigma, \sigma,  \\ \alpha<\beta}}   d_{i\alpha\sigma}^{\dagger} d_{i\beta\sigma'}^{\dagger} d_{i\alpha\sigma'}^{\phantom{\dagger}} d_{i\beta\sigma}^{\phantom{\dagger}}+ J_P \sum_{\substack{i,  \sigma, \\\alpha<\beta }} d_{i\alpha\sigma}^{\dagger}
d_{i\alpha\bar{\sigma}}^{\dagger} d_{i\beta\bar{\sigma}}^{\phantom{\dagger}} d_{i\beta\sigma}^{\phantom{\dagger}} 
\end{split}
\end{equation*}
Here, $U$ is the intraband Hubbard repulsion, $U'$, $J_H$, and $J_P$ are interband Hubbard repulsion, Hund's exchange interaction, and pair hopping amplitude, respectively. 


The parameters for this model for modeling nickelate and cuprates have been extensively reported in the literature and notably varied. In  [supplement~\cite{SM} Table~I], we present the adapted non-interacting parameters from a few studies~\cite{PhysRevX.10.011024, PhysRevX.10.021061, Hepting2020, PhysRevLett.126.127401}. The literature has a notably varied set of parameters, albeit all these studies agree that nickelates have large TM-ligand spitting and larger hopping overlap ($t_{pd}$) compared to cuprates. 
The relevant non-interacting parameters for this work are also shown in [supplement~\cite{SM} Table~I]. 
For the interacting part, we use $U_{d}=6.5$, $U_{p} =4.1,  V_{pd} = 1.0$ and $J_H=0.8$ (unless otherwise stated) for the nickelate model. 


For evaluating the RIXS spectra, we consider the core-hole Hamiltonian given by 
$H_\text{ch} = \mathcal{H} + U_\text{ch}\sum_{i} n_i^d n_i^p$ for the $L$-edge~\cite{Nocera2018, UKumar2019}. 
RIXS cross-section is evaluated using the Kramers-Heisenberg formula~\cite{RevModPhys.83.705, Kumar_2018,Schlappa2018} given by, 
\begin{equation}\label{eq:KH}
\begin{split}
\mathcal{I} =  \sum_f \bigg| \sum_{i, n, \sigma} \frac{\langle f|\mathcal{D}^\dagger| n\rangle\langle n| \mathcal{D} |g\rangle}{E_n\minus E_g\minus\hbar\omega_\text{in} +\iota\Gamma_n} \bigg|^2  \delta(E_f \minus E_g \minus \Omega).
\end{split}
\end{equation}
Here, $|g\rangle$, $|n\rangle$ and $|f\rangle$ are the ground, intermediate, and final states with energies $E_g$, $E_n$, and $E_f$ respectively and $\Omega~(= \hbar\omega_\text{in}-\hbar\omega_\text{out})$ is the energy loss of the incoming photon. $\Gamma$ is the inverse core-hole lifetime.  $\mathcal{D}$ is dipole operator given by 
$\mathcal{D}$ = $\sum_{i,\alpha,\sigma, J} e^{i{\bf k}_i\cdot {\bf R}_i} p_{i,J}^\dagger d_{i,\alpha,\sigma}^\pdag $  for the Ni $L$-edge and involve core $2p$ (with spin-orbit coupling) and valence $3d$ shells of the nickel site.  We evaluate spin-resolved Ni $L$-edge RIXS, where the angular dependence is integrated out, and the spectra are resolved into non-spin conserving (NSC, $\Delta S= 1$) and spin conserving (SC, $\Delta S= 0$) channels, depending on odd or even number of spin-flips in the excitations~\cite{PhysRevB.85.064423,PhysRevX.6.021020, UKumar2019, UKumar2022, MorettiSala2011}. 

{\sl Results\label{results}:---}  We now present  results for the nickelates using a combination of techniques. The nickelate model allows for interesting scenarios, in particular for the doped case, as the doped nickelates model has larger occupancies on $d$-orbitals for the case of doped nickelates (also see supplement Fig. S2~\cite{SM}). We present nickelates  RIXS spectra and investigate the origin of the softening in the RIXS spectra.  

Fig.~\ref{fig:Schematicsdd} shows the cluster considered in our work and schematics for the $dd$ excitations. Panel (a) shows the Ni$_2$O$_7$ cluster considered later in our work, and panel (b) shows the energy Ni $3d$ onsite levels considered in our cluster. 
The Ni cluster with ligand can be mapped to energy levels for the single Ni site, integrating out the ligand (see supplement Sec. S2 for the mapping). We, therefore, start by considering the different scenarios of $dd$ excitation for a single Ni ion (undoped: $d^9$ and doped: $d^8$) case. We consider that the lowest energy orbital is $d_{x^2-y^2}$, and other $d$ orbitals are roughly separated by $\Delta$ energy gap in the hole language.

Panel c) presents the schematics for the $dd$ excitations realized in undoped single-site nickelates, where $dd$-excitations can be understood as simple excitations between $d_{x^2-y^2}$ to other $d$ orbitals ($\Delta$) as elucidated in the panel. In the doped case, there are two possible ground state configurations depending on the strength of $J_H$ and $\Delta$: i) $\Delta>J_H$: doubly occupied $d_{x^2-y^2}$, or ii) large $J_H>\Delta$:  one hole each in $d_{x^2-y^2}$ and $d_{z^2}$ (the preferential occupancy on this orbital is discussed later) as shown in panel d) and e) respectively. The NSC channel in d) leads to the softening of $dd$ excitations. This configuration is also consistent with the ground state occupancy estimates.  On the other hand,  e) ground state will lead to hardening of the $dd$ excitations. We use these schematic pictures as a guiding tool to understand the spectra in the models below. 

\begin{figure}
\centering
\includegraphics[width=1.0\linewidth]{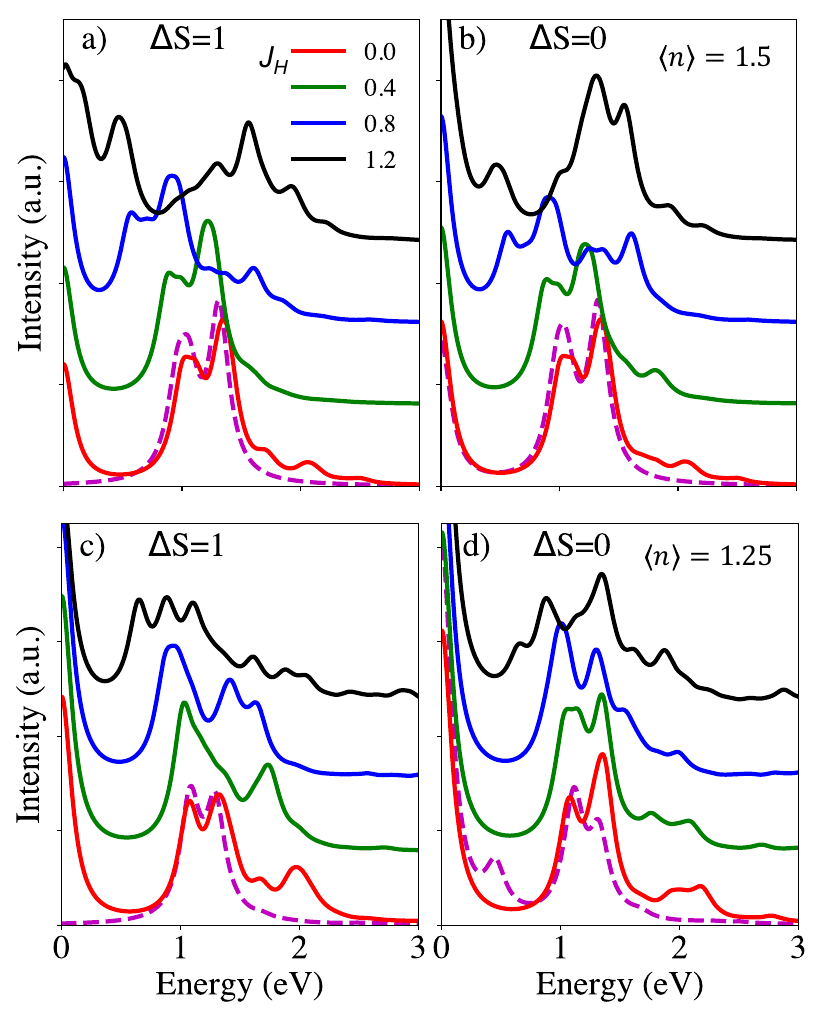}
\vspace{-0.5cm}
\caption{Spin-resolved Ni $L$-edge RIXS spectra. a) and b) show results for Ni$_2$O$_7$ cluster in the NSC $(\Delta S=1)$ and SC $(\Delta S=0)$ channels. The dashed magenta line shows spectra for the undoped ($\langle n\rangle =1$)  model. Solid lines show spectra for the doped ($\langle n\rangle =1.5$) model for the values of $J_H$ mentioned in the legend. 
Similarly, c) and d) show results for Ni$_4$O$_8$ cluster in the NSC and SC channels, respectively, for undoped and doped ($\langle n\rangle =1.25$) models.\label{fig:RIXSLedge}}
\vspace{-0.5cm}
\end{figure}

{\sl RIXS spectra:}  We present the simulated RIXS spectra for a set of models to consider realistic doping to decipher the effect of $J_H$. 
We use Ni$_2$O$_7$ cluster to explore the undoped and hole-doped (50\%) model. We additionally present results for a smaller doping (25\%) model using Ni$_4$O$_8$ cluster and Anderson impurity model. 
To make Ni$_4$O$_8$ cluster computationally accessible, we neglect $d_{yz/zx}$ orbitals in the cluster [for details, see supplement~\cite{SM}]. Using these models, we examine the role of $J_H$ in the spectra.
 
 We start by examining the Ni $L$-edge RIXS spectra using Ni$_2$O$_7$ cluster with ($1\uparrow,1 \downarrow$) and three ($2\uparrow,1 \downarrow$) holes fillings to simulate the undoped and doped (50\%) models in both the non spin-conserving (NSC, $\Delta S=1$) and spin conserving (SC, $\Delta S = 0$) channels using KH formalism given by Eq.~\ref{eq:KH}.

\begin{figure}
\centering
\includegraphics[width=0.9\linewidth]{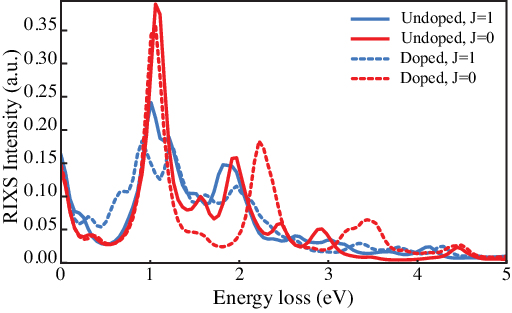} 
\vspace{-0.25cm}
\caption{RIXS spectra from the AIM for both the undoped and doped (0.25 hole doping) model. The results with and without $J_H$ are shown in blue and red, respectively.\label{fig:EDRIXS_AIM}}
\end{figure}

Panels a) and b) in Fig.~\ref{fig:RIXSLedge} show NSC and SC channels of the Ni $L$-edge for the Ni$_2$O$_7$ cluster. The dashed magenta line shows the RIXS spectra for the undoped model and is independent of $J_H$.  The doped (50\%) model results are also shown in both panels using solid lines for a set of the $J_H$. The prominent peaks for the doped model $J_H = 0$ occur at the same energies as in the undoped case. We notice a softening of $dd$-excitation in both channels with the increase in $J_H$ in both the NSC and SC channels. We will further revisit these results from this cluster later to examine the nature of the various peaks in both the channels.

To examine results for the lower doping (25\%), we consider Ni$_4$O$_8$ cluster with ($3\uparrow, 2\downarrow$) filling and evaluate the Ni $L$-edge RIXS spectra. Panels c) and d)  case show results for both the undoped and doped models for this cluster. The dashed magenta line shows the RIXS spectra for the undoped model and is independent of $J_H$. In the SC channel, an additional bimagnon peak at around 0.3 eV~\cite{PhysRevB.85.214527,Pal2023, Schlappa2018} is observed for the undoped case. The large value of bigmagnon can be attributed to the finite size of the cluster.  The doped (25\%) model results are shown using solid lines. The $J_H = 0$ case has the same characteristics as the undoped case. We observe softening of $dd$-excitation in both cases with an increase in $J_H$. We, therefore, observed softening here for realistic doping. 

These cluster results reveal the softening of $dd$ excitations on doping consistent with reported RIXS experiment on NdNiO$_2$~\cite{PhysRevB.104.L220505} and more recently in trilayer nickelates where the $dd$ excitations were reported at even lower energies~\cite{YShen2023}. This shows that Hund's coupling is important for understanding the softening of $dd$ excitations observed in nickelates.

\begin{figure}
\centering
\includegraphics[width=1\linewidth]{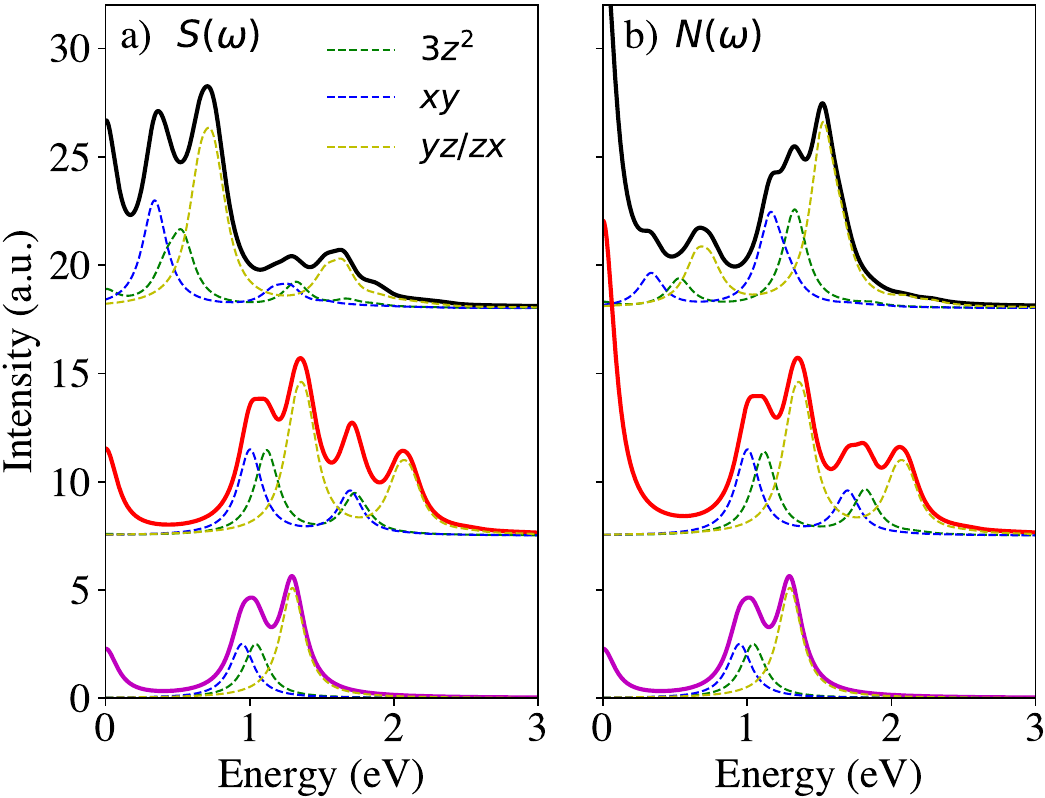}
\vspace{-0.25cm}
\caption{ $S(\omega)$ and $N(\omega)$ for the Ni$_2$O$_7$ cluster are shown in panel a) and b). Solid magenta shows the response for the undoped model. The dashed lines show the orbital-resolved excitations in the spectra. Solid red and black lines show response for the doped model with $J_H= 0$ and 1, respectively. \label{fig:SwNw}}
\vspace{-0.5cm}
\end{figure}

We now present the RIXS spectra using an Anderson Impurity model (AIM) [see supplement S6~\cite{SM} and Ref.~\cite{WANG2019151} for details]  derived within charge self-consistent DFT+DMFT using  Portobello~\cite{PhysRevB.85.155129, melnick_accelerated_2021, adler_portobello_2023}. The DFT+DMFT calculations are conducted with $U=5$ eV and $J_H=1$ and 0 eV using the GGA functional~\cite{perdew_generalized_1996} and fully-localized limit double counting with a nominal Ni-$d^8$ valence. 
Fig.~\ref{fig:EDRIXS_AIM} shows the Ni $L$-edge RIXS spectra for the AIM. The undoped results are shown in solid lines using red and blue colors for $J_H = 0$ and $J_H=1$, respectively. The $dd$-excitations exist at 1-2.5 eV and are minimally affected by $J_H$.  The RIXS spectra for the doped model  are shown in the dashed line in the respective colors. We find a significant softening of $dd$-excitations for the $J_H =1$ case compared to the undoped case and the $J_H=0$ of the doped case. This is consistent with our small cluster calculations and confirms the softening of $dd$-excitations observed in the experiment~\cite{PhysRevB.104.L220505}.

{\sl Dynamical spin and charge response:---} 
The RIXS response given by Eq.~\ref{eq:KH} is complex; we, therefore, examine the model in the $\Gamma_n \rightarrow \infty$ limit, in which the core-hole effects are eliminated. We evaluate the dynamical spin ($S(\omega)$, $\Delta S=1$) and charge ($N(\omega)$, $\Delta S=0$) response on the Ni$_2$O$_7$ cluster.  Fig.~\ref{fig:SwNw} shows the $S(\omega)$ and $N(\omega)$ response in the panel a) and b), respectively. The undoped model results are shown in magenta solid line for $J_H=1.2$ and the doped model in red and black lines for $J_H =0$ and 1, respectively. We also plot the orbital resolved spectra where the outgoing dipole operator in Eq.~\ref{eq:KH} is restricted to a single orbital, which allows us to identify the different $dd$-excitations. 
In both spectra, the ordering of the orbital excitations with energy less is as follows for the undoped model: $d_{xy}$, $d_{3z^2}$, and then $d_{yz/zx}$. This hierarchy of $dd$ excitations differs from Ref.~\cite{PhysRevB.104.L220505} which reports that $3z^2$ is the highest excitations in NdNiO$_2$ using the polarization dependence of the excitations. We re-examine the data in [see supplement S5~\cite{SM}] and show that assigning $d_{3z^2}$ the lower energy excitations compared to $d_{yz/zx}$ orbital excitation can also reproduce the observed polarization dependence, as these orbitals have similar angular dependence within the reported experimental resolutions.   

 In the doped model, the results are sensitive to $J_H$. For $J_H = 0$, the orbital resolved $dd$-excitations has primarily two peaks, with the bright lower energy peaks coinciding with the undoped case. For $J_H= 1$, we find that the prominent excitations soften in the $S(\omega)$ channel, consistent with the RIXS results. 
This contrasts with the $N(\omega)$,  where mild hardening is observed, and the softening in SC channel of RIXS can be described as a concequence of finite core-hole lifetime.

\begin{figure}[t]
\centering
\hspace{-0.2cm} 
\includegraphics[width=1.0\linewidth]{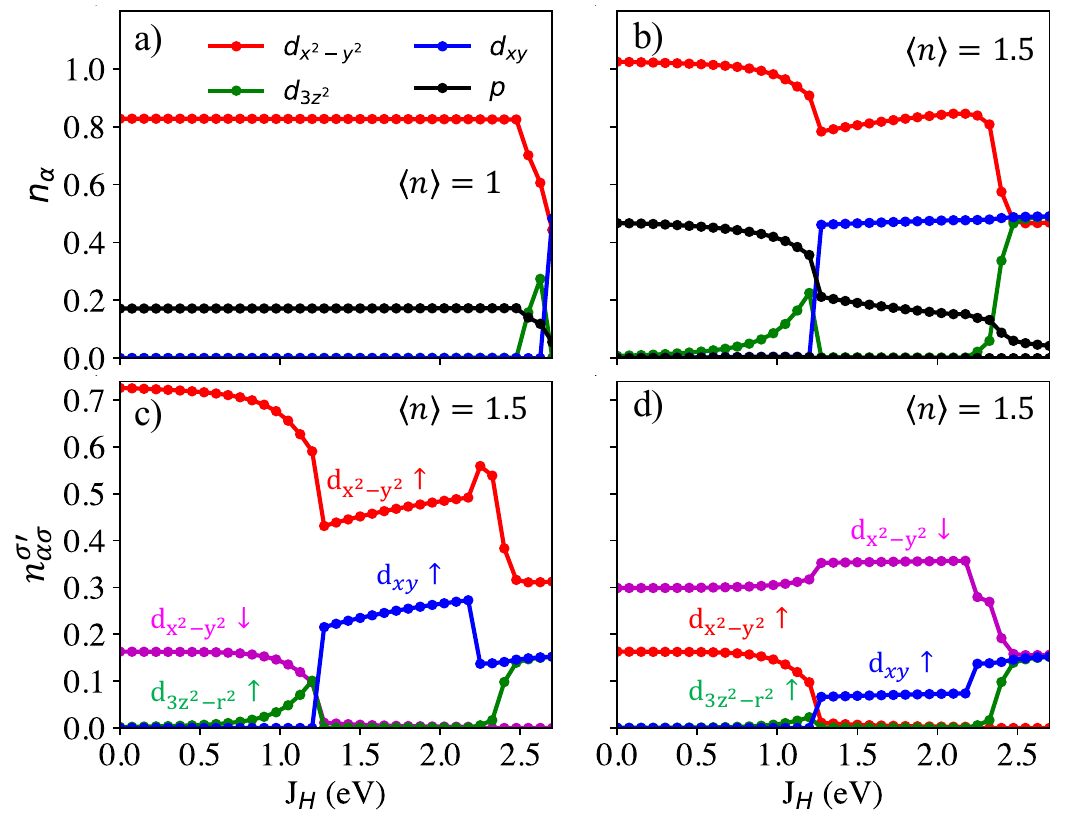}
\vspace{-0.25cm} 
\caption{Ground state analysis of the Ni$_2$O$_7$ cluster. a) and b) show the orbital resolved $n_{\alpha}$ dependence on $J_H$ for undoped ($\langle n\rangle =1$) and doped ($\langle n\rangle =1.5$), respectively. c) and d) show the spin-resolved orbital double occupancies (${n}_{\alpha\sigma}[n_{d_{x^2-y^2}\sigma'}] $) on the Ni site in the doped case for $\sigma' = \uparrow$ and  $\downarrow$, respectively. }
\vspace{-0.5cm} 
\label{fig:Ni2O7GS}
\end{figure}

Fig.~\ref{fig:Ni2O7GS} shows the results for the analysis of the ground state dependence on $J_H$ in the undoped ($\langle n\rangle =1$) ) and doped ($\langle n\rangle =1.5$).  Panel a)  shows results for the undoped model. Here, the occupancies in the Ni $3d$ orbitals ($n_\alpha$) in the ground state, which is given by $n_\alpha = \langle g|\sum_\sigma  n_{i,\alpha,\sigma}|g\rangle$ dependence on $J_H$ are presented. We find that the holes are restricted to $d_{x^2-y^2}$ orbital and the oxygen $p$ orbitals. Other $d$-orbitals do not contribute to the ground state for reasonable $J_H$. 

Panels b)-d) shows the ground state results for the doped model. Panel b) shows $n_\alpha$ dependence on $J_H$. For small $J_H$, the has holes restricted to the $d_{x^2-y^2}$ orbital and O $2p$ orbitals, the occupancy on $d_{3z^2}$ increases with $J_H$ till 1.25 eV, where it reaches around 20\% of $d_{x^2-y^2}$. This is despite $d_{xy}$ having a lower onsite energy. The dominant weight on $d_{3z^2}$ can be attributed to the large hopping and the oxygen $p$-orbitals shared between the two $d$-orbitals (also see Fig.~S1 in supplement~\cite{SM}).  For $J_H=1.25-2.25$ eV, the $d_{xy}$ orbital becomes occupied with 60\% the occupancy of the $d_{x^2-y^2}$ orbital.  For the unphysical $J_H>2.5$ eV, these three orbitals are almost equally occupied.
To investigate where the extra hole resides, we plot the constrained occupancies,  $n_{\alpha\sigma}^{\sigma'} = \langle g|n_{d_{\alpha\sigma}} [n_{d_{x^2-y^2\sigma'}}]|g \rangle$. The occupancy on $d_{x^2-y^2}$ with spin $\sigma' = \uparrow$ and $\downarrow$ is constrained and are shown in panels c) and d), respectively. For small $J_H$, the $d_{x^2-y^2}$  has double occupancy, which is expected due to the penalty of the CF-splitting. We notice an increase in occupancy on $d_{3z^2}$ with $\sigma= \sigma'$ mediated by Hund's coupling till $J_H = 1.25$ eV. This makes the scenario discussed in Fig.~\ref{fig:Schematicsdd}(d) relevant for the nickelates, which leads to the softening of $dd$ excitations. 

{\sl Conclusions\label{conclusions}:---} In this work, we have explored the effects of Hund's coupling in undoped and doped nickelates using small cluster models and DMFT. Using these models, we find that Hund's coupling plays a central role in softening the $dd$ excitations and softening is enhanced for larger doping. This softening of $dd$-excitations with the increase in doping is consistent with the recent report experimental study of IL nickelates~\cite{PhysRevB.104.L220505}. Our small cluster analysis shows that doubly occupied $d_{x^2-y^2}$ is the preferred state for the doped nickelates, but the $dd$ excitations are softened due to lowering of the $dd$ excitations in the RIXS spectra. This contrasts with the cuprate, where the extra hole typically resides on the ligands~\cite{PhysRevB.37.3759}. 

Future experiments using photon polarization dependent RIXS that allow for spin resolution~\cite{PhysRevX.12.021041}.   They will test  the relative contributions of  spin and charge fluctuations to the RIXS signal.  Comparison with simulations will   further help one clarify the minimal model for the nickelates. Also, a comparative study of $dd$ excitations nickelates and cuprates can be a useful tool to highlight the difference between the two.


This work was supported by the US Department of Energy, Office of Basic Energy Sciences, as part of the Computation Material Science Program.

\bibliography{refNi}

\makeatletter

\pagebreak
\widetext
\begin{center}
	\textbf{\large Supplement for ``Softening of $dd$ excitation in the resonant inelastic x-ray scattering spectra as a signature of Hund's coupling in nickelates"}
\end{center}
\setcounter{equation}{0}
\setcounter{figure}{0}
\setcounter{table}{0}
\setcounter{page}{1}
\makeatletter

\renewcommand{\theequation}{S\arabic{equation}}
\renewcommand{\thefigure}{S\arabic{figure}}
\bibliographystyle{apsrev4-1}
\setcounter{section}{0}
\renewcommand{\thesection}{S\arabic{section}}
	



\section{Parameters for the model Hamiltonian}

A number of studies have reported relevant parameters for the nickelates and cuprates. In Table~\ref{Table:LitParameters}, we present the adapted parameters from a few studies. The literature has a notably varied set of parameters. Still, all these studies report smaller $3d$-orbitals crystal field splitting compared to the cuprates. Additionally, larger hopping overlap ($t_{pd}$) and oxygen $p$ onsite energies ($\Delta_{pd}$) are reported for nickelates. 
\begin{table}[h]
\begin{tabular}{ll|lllll} \\
\hline
 &  & $\epsilon_{3z^2}$~~ & $\epsilon_{xy}$~~ &$\epsilon_{yz/zx}$~ & $\Delta_{pd}$  & $t_{pd}$~~~  \\ \hline
Ref.~\cite{PhysRevX.10.011024} & LaNiO$_2$ & 0.71 & 0.73 & 0.63 & 4.4 & 1.23 \\ 
  & CaCuO$_2$ & 0.97 & 1.04 & 0.93 & 2.69 & 1.20 \\ 
Ref.~\cite{PhysRevX.10.021061} & NdNiO$_2$ & 0.14 & - & - & 3.3 & 1.37 \\ 
 & CaCuO$_2$ & 0.56 & - & - & 2.1 & 1.27 \\ 
Ref.~\cite{YShen2022} & La$_4$Ni$_3$O$_8$ & 0.2 & 0.1 & 0.3 & 5.6 & 1.36 \\
  & La$_{2\minus x}$Sr$_x$CuO$_4$ & 0.95 & 0.7 & 0.9 & 4.0 &  1.17\\
Ref.~\cite{Hepting2020} & LaNiO$_2$ & 0.74 & 0.8 & 0.70 & 3.96 &  1.20\\
Ref.~\cite{PhysRevLett.126.127401} & LaNiO$_2$ & 0.38 & 2.03 & 1.40 & - & - \\ 
This work & Ni cluster & 0.2 & 0.1  & 0.3 &  5.6 & 1.36 \\
& Cu cluster & 0.95 & 0.7  & 0.9 &  4.0 & 1.17 \\
\hline 
\end{tabular}
\caption{Parameters reported in the literature for nickelates and their comparison with cuprates. The onsite energies are with respect to the $\epsilon_{x^2-y^2}$ as a reference in the hole language. $\Delta_{pd}$ is the onsite energy of oxygen $p$-orbitals with respect to the $\epsilon_{x^2-y^2}$. $t_\text{pd}$  is the overlap of $d_{x^2-y^2}$ and oxygen $p$ orbitals.\label{Table:LitParameters}}
\end{table} 

In this work, we use the parameters shown in Table~\ref{Table:LitParameters}, motivated by Ref.~\cite{YShen2022}. We use the relation; $V_{pd\pi} = -V_{pd\sigma}/2$ and $V_{pp\pi} = -V_{pp\sigma}/4$ to evaluate the hopping parameters between other orbitals shown in Fig.~\ref{fig:OrbitalsOverlap}. We also use; $U_{d}=6.5$, $U_{p} =4.1$ and $J_H=0.8$ (unless otherwise stated) for the nickelates.  These parameters are consistent with the observed superexchange in TL nickelates~\cite{JQLin2021} ($i.e.~ 69$ meV) and IL nickelates  ($i.e. ~64$ meV)~\cite{HLu2021,Gao2023}. As $J \approx 4 \frac{t_{pd}^4}{\Delta_{pd}^2 U_d}$, the large value of $t_{pd}$ allows for a significant superexchange in spite of the large $\Delta_{pd}$~\cite{UKumar2021, PhysRevLett.124.207004}.


\section{Mapping TMO$_4$ cluster to one-site cluster}\label{App:TMO4toOnesite}
\begin{figure}
\centering
\includegraphics[width=0.5\linewidth]{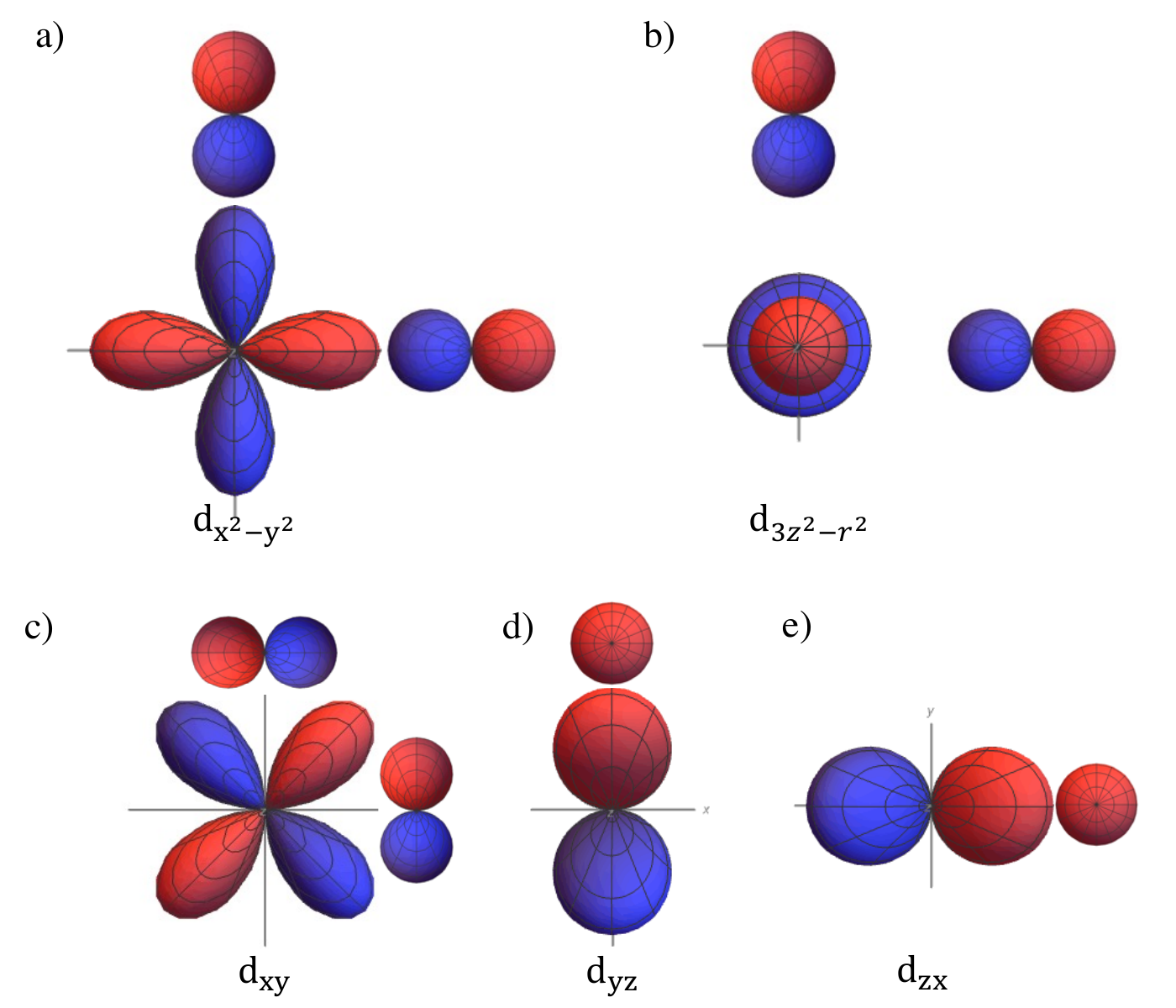}
\caption{ Top view of the hopping overlaps for the $d$ orbitals with the adjacent oxygen orbitals.  a) and b) show $d_{x^2-y^2}$ and $d_{3z^2}$ overlap with the same oxygen $p$-orbitals, in contrast to other $d$-orbitals shown in c)-e), which hybridize with the other oxygen $p$-orbitals.}
\label{fig:OrbitalsOverlap}
\end{figure}

In the TMO$_4$ plaquettes at half-filling (one hole in the cluster), hybridized energies of states for $d_{x^2-y^2}$ orbitals (see Fig.~\ref{fig:OrbitalsOverlap}(a)) can be estimated using 
\begin{equation}
	H= {\bf C}^\dagger \left(
	\begin{array}{ccccc}
		0 & -t_{\text{pd}} & t_{\text{pd}} & t_{\text{pd}} & -t_{\text{pd}} \\
		-t_{\text{pd}} & \Delta_{pd}  & t_{\text{pp}} & 0 & -t_{\text{pp}} \\
		t_{\text{pd}} & t_{\text{pp}} & \Delta_{pd}  & -t_{\text{pp}} & 0 \\
		t_{\text{pd}} & 0 & -t_{\text{pp}} & \Delta_{pd}  & t_{\text{pp}} \\
		-t_{\text{pd}} & -t_{\text{pp}} & 0 & t_{\text{pp}} & \Delta_{pd}  \\
	\end{array}
	\right)  {\bf C} 
\end{equation}
Here, ${\bf C}^\dagger = \begin{pmatrix} d_{\alpha}^\dagger & p_{\alpha, +x}^\dagger & p_{\alpha, {+y}}^\dagger & p_{\alpha, \minus x}^\dagger& p_{\alpha, \minus y}^\dagger\end{pmatrix}$. Similarly, one can write the Hamiltonian associated with other $d$-orbitals. The lowest eigen-energies for the $d$-orbitals are given by 
\begin{equation}
	\begin{split}
		&E_{x^2-y^2}  =  \frac{1}{2} \left( \tilde{\Delta}_{-} -\sqrt{16 t_{\text{pd}}^2+ {\tilde{\Delta}_{-}}^2}\right) \\
		& E_{3z^2/xy} = \frac{1}{2} \left({\tilde{\Delta}_{+}}+\epsilon_\alpha-\sqrt{16 T_\alpha^2+\big({\tilde{\Delta}_{+}}-\epsilon_\alpha \big)^2} \right)  \\
		& E_{yz/zx} = \frac{1}{2} \left(\Delta +\epsilon_\alpha -\sqrt{8T_\alpha^2+(\Delta -\epsilon_\alpha )^2} \right) \\
	\end{split}
\end{equation}
Here $\tilde{\Delta}_\pm= \Delta_{pd} \pm 2 t_{\text{pp}}$ and $\epsilon_\alpha$ and $T_\alpha$ are the respective on-site energy and hopping integral for the respective $d$-orbitals.  Notice that the oxygen onsite energies are renormalized differently for $x^2-y^2$ orbitals and the  $3z^2/xy$ orbitals due to distinct phases in these orbitals.  The $\Delta_{pd}$ for $yz/zx$ is unaffected as we consider hopping only in th $xy$-plane, see Fig.~\ref{fig:OrbitalsOverlap}(d)-(e).  Using the parameters in Table~\ref{Table:LitParameters}, the dd-excitations therefore associated with various excitations, in the limit, $\{t_{pd}, t_{pp}, \epsilon\} \ll \Delta_{pd} $, can be approximate to  

\begin{equation}
\begin{split}
& \Delta E_{3z^2} \approx  \epsilon_{3z^2} + \frac{4t_\text{pd}^2}{3\Delta_{pd}}\left(2+\frac{8t_\text{pp}}{\Delta_{pd}} -\frac{\epsilon_{3z^2}}{\Delta_{pd}}\right) \\
& \Delta E_{xy}\approx  \epsilon_{xy} + \frac{4t_\text{pd}^2}{4\Delta_{pd}}\left(3+\frac{10t_\text{pp}}{\Delta_{pd}} -\frac{\epsilon_{xy}}{\Delta_{pd}}\right)  \\
& \Delta E_{yz/zx}  \approx  \epsilon_{yz/zx}  + \frac{4t_\text{pd}^2}{8\Delta_{pd}}\left(7+\frac{16t_\text{pp}}{\Delta_{pd}} -\frac{\epsilon_{yz/zx} }{\Delta_{pd}}\right)
\end{split}
\end{equation}
These energy differences give an estimate for the dd excitations for the undoped model.  Notice that the hybridized $dd$-excitation depends on the onsite energies. Also, the $t_\text{pd}^2/\Delta_{pd}$ is significant and can, therefore,  strongly renormalize the $dd$-excitations.

\section{Ground state analysis\label{App:GSAnalysis}}

\begin{figure}[h]
\centering
\includegraphics[width=\linewidth]{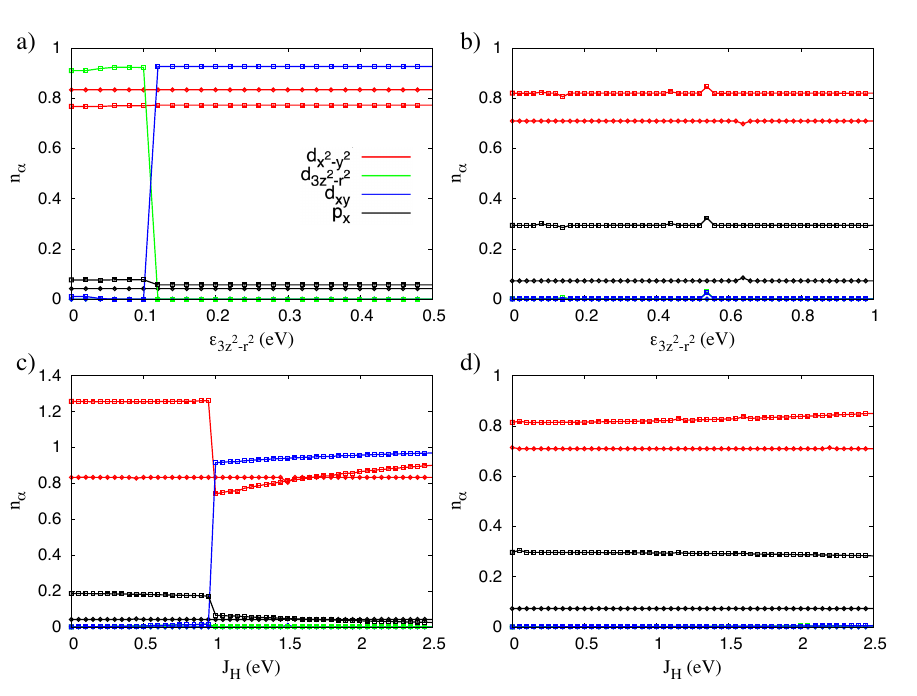}
\caption{Orbitals occupancy for the one ($i.e.~d^9$) and two ($i.e.~d^8$) holes  in the TMO$_4$ clusters. a) and c) show the orbital-resolved occupancy results for NiO$_4$ cluster dependence on tetragonal distortion, characterized by $\epsilon_{3z^2}$ and $J_H$, respectively. b) and d)  show the orbital-resolved occupancy results for the CuO$_4$ cluster. Open circles ($\circ$) and squares ($\square$) plot the results for one and two holes, respectively.}
\label{fig:occupancynio4d8}
\end{figure}



To understand our model in a simplistic case with all the interactions,  we explore the TMO$_4$ cluster with $3d$ of TM (transition metal) and $2p$ oxygen orbitals in the model. 
We highlight the differences between the model for nickelate and cuprates. The $d^9$ and $d^8$ configurations suggest that the multiorbital effects are important for nickelates compared to the models used for cuprates. 

{\bf TMO$_4$ cluster:---} Fig~\ref{fig:occupancynio4d8} shows the ground state orbital resolved occupancy results for one and two holes in NiO$_4$ cluster and CuO$_4$ clusters. 
Panel a) shows orbital-resolved occupancy results for NiO$_4$ cluster dependence on the onsite energy of $\epsilon_{3z^2}$ for NiO$_4$  cluster for one and two holes, respectively. Similarly, panel b) shows the results for the CuO$_4$ cluster.
We notice that the ground state for the one hole is restricted to the $d_{x^2-y^2}$ orbital in both the case and hybridized with the ligands. This is because the Coulomb terms cannot act on a single hole. In the case of two holes, the NiO$_4$ plotted for $J_H = 1.2$ has the additional hole residing on the $d_{3z^-r^2}$ till $\epsilon_{3z^2}\approx \epsilon_{xy}$ after that the hole instead resides on $d_{xy}$. In contrast, the extra hole resides on the ligand in CuO$_4$ model due to smaller onsite ligand energies.  The milder dependence increase in oxygen site occupancy in nickelates is supported by the electron energy loss spectroscopy (EELS) on oxygen $K$-edge of nickelates~\cite{Goodge2021, Goodge2023} with doping in contrast to the large increase in oxygen occupancy of hole-doped cuprates~\cite{PhysRevB.96.115148}.

Panel c) and d) show the orbital-resolved occupancy results for NiO$_4$ and CuO$_4$ clusters, respectively, dependence on $J_H$. Again in the single hole case, the hole is restricted to $d_{x^2-y^2}$-orbital and hybridized with ligands for both clusters. In the case of two holes, we notice that for small $J_H< 1$ in NiO$_4$ cluster, the extra hole resides on the $d_{x^2-y^2}$. Beyond this, the extra hole resides on the $d_{xy}$ orbital and the relevant oxygen orbitals. Instead for the CuO$_4$ cluster, the extra hole always resides on the ligand side, consistent with the charge-transfer picture of cuprates. 

While the single site model can only account for the local effects, our model shows that the single $d$ orbital picture works for both the nickelate and cuprate in the undoped model. In the doped model, multiple orbitals can play a significant role for nickelate, in contrast to cuprates where occupancy is still restricted to $d_{x^2-y^2}$ along with the extra hole preferentially residing on the ligand. \\

\begin{figure}[t]
	\centering
	\hspace*{-0.25cm} 
	\includegraphics[width=0.7\linewidth]{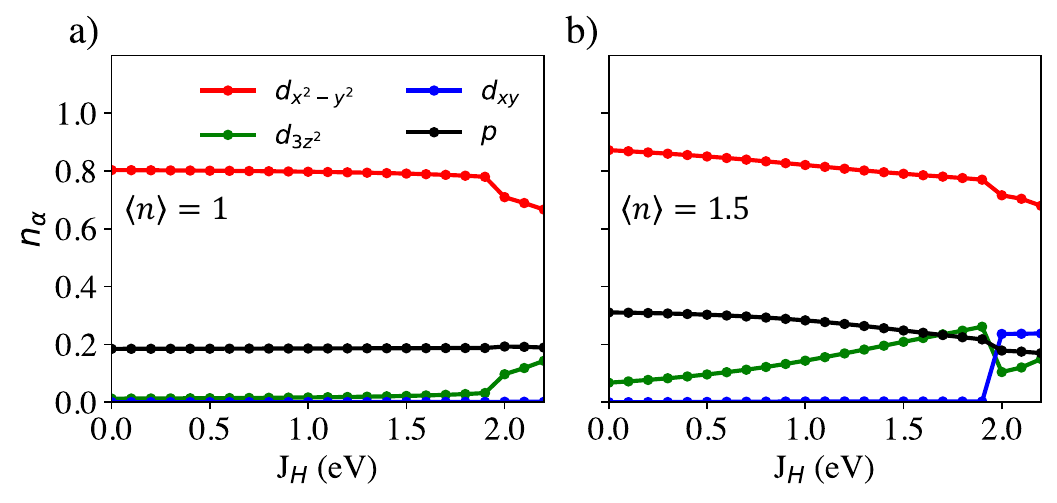}
	\caption{Orbital resolved occupancies in the ground state of Ni$_4$O$_8$ cluster. a) shows the dependence of orbital-resolved occupancies on $J_H$ for the undoped cluster, and b) shows the dependence of orbital-resolved occupancies on $J_H$  for the doped (25\%) cluster.}
	\label{fig:ni4o8hundsoccupancy}
\end{figure}

We have reported the results for the Ni$_2$O$_7$ cluster in the main text. We present results for Ni$_4$O$_8$ cluster without $d_{yz/zx}$ orbitals in the cluster to make it computationally accessible. 

{\bf Ni$_4$O$_8$ cluster:---} Here, we report the results for the state-of-the-art Ni$_4$O$_8$ cluster, which allows for 25\% doping ($d^{8.75}$) in the model Hamiltonian cluster. To make this computationally accessible, we integrate out the $d_{yz/zx}$-orbitals as these do not contribute to the ground state, as is evident from the smaller cluster work. The reduced Hilbert space, therefore, allows us to explore 1.25 hole/site $(i.e.~d^{8.75})$ in the cluster, which is relatively close to the superconducting IL-nickelates~\cite{PhysRevLett.125.027001}. We use periodic boundary conditions for this model.

Fig.~\ref{fig:ni4o8hundsoccupancy} shows the $n_\alpha$ dependence on $J_H$ in the ground state for this cluster.  Panel a) shows results for the undoped model. The ground state occupancies are unaffected by Hund's coupling for reasonable values of $J_H~(<1.8)$, consistent with the other clusters. The holes are hybridized between $d_{x^2-y^2}$ and the oxygen orbitals. We notice the shift of weight from $d_{x^2-y^2}$ to $d_{3z^2}$ for $J_H>1.8$, as for this large value, the other orbital becomes competitive. 
Panel b) shows the result for 25\% hole-doped ($i.e. ~d^{8.75}$) model. The extra hole is distributed between oxygen and $d_{x^2-y^2}$ orbital for small $J_H$.  With increasing $J_H$, the hole weight in $d_{3z^2}$ keep increasing. In fact, for reasonable $J_H$, the $d_{3z^2}$ can be almost a quarter of the weight of $d_{x^2-y^2}$, leading to a significant contribution to the ground state. This suggests that Hund coupling enhances multiorbital effects even for small doping.

\begin{figure}
\centering
\hspace*{-4mm}
\includegraphics[width=0.7\linewidth]{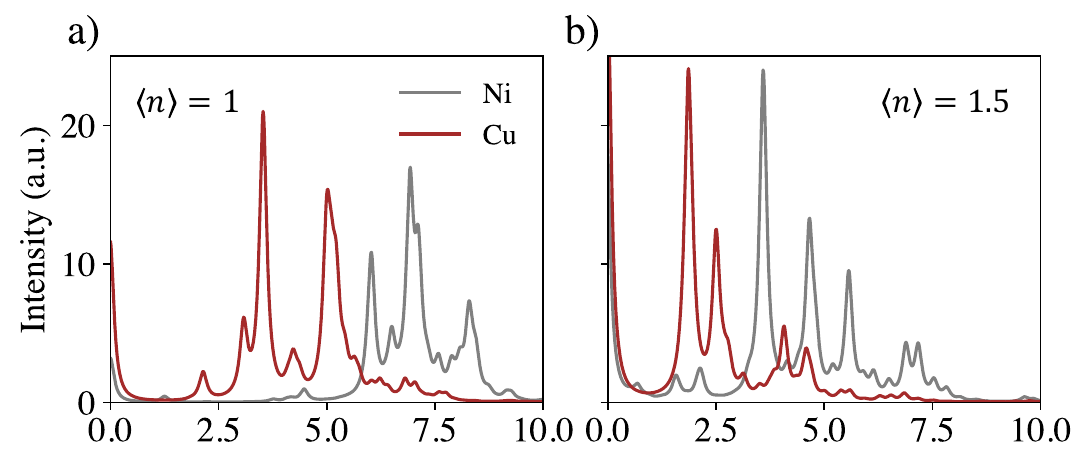}
\caption{Oxygen $K$-edge RIXS spectra for TM(=Ni, Cu)$_2$O$_7$ cluster. a) and b) show the spectra for the undoped and doped models, respectively. The gray (brown) line in each panel shows the spectra for nickelate (cuprates) models.\label{fig:rixsni2o7_OKedge}}
\end{figure}

\section{Oxygen $K$-edge RIXS}
We, here, present the O $K$-edge RIXS spectra using the TM$_2$O$_7$ clusters for nickelates and cuprates to explore the charge transfer excitations in these. We contrast the spectra of the nickelates with the cuprates. 
For evaluating the RIXS spectra, we consider the core-hole Hamiltonian given by $H_\text{ch} = \mathcal{H} + U_\text{ch}\sum_{i} n_i^p n_i^s$ for the oxygen $K$-edge~\cite{YShen2022, Schlappa2018, Kumar_2018}. 
RIXS cross-section is evaluated using the Kramers-Heisenberg formula given by, 
\begin{equation}
	\begin{split}
	\mathcal{I}_{RIXS} =  \sum_f \bigg| \sum_{i, n, \sigma} \frac{\langle f|\mathcal{D}^\dagger| n\rangle\langle n| \mathcal{D} |g\rangle}{E_n - E_g-\hbar\omega_\text{in} +\iota\Gamma_n} \bigg|^2 \\ \times \delta(E_f - E_g -\Omega).
	\end{split}
\end{equation}
Here, $|g\rangle$, $|n\rangle$ and $|f\rangle$ are the ground, intermediate, and final states with energies $E_g$, $E_n$, and $E_f$ respectively and $\Omega~(= \hbar\omega_\text{in}-\hbar\omega_\text{out})$ is the energy loss of the incoming photon. $\Gamma$ is the inverse core-hole lifetime.  $\mathcal{D}$ is dipole operator given by $\mathcal{D}$ = $\sum_{i,\alpha, \sigma} e^{i{\bf k}_i\cdot {\bf R}_i} s_{i,\sigma}^\dagger p_{i,\alpha,\sigma}^\pdag  $ for oxygen $K$-edge which involves the core $1s$ and valence $2p$ shells of the oxygen sites~\cite{YShen2022}. 

Fig.~\ref{fig:rixsni2o7_OKedge} shows the O $K$-edge spectra of the nickelate and cuprate. 
Panel a) shows O $K$-edge RIXS spectra for the undoped model. The brightest feature for the nickelate appears at around 5.5-8 eV, whereas in cuprates, it appears at around 3-5 eV~\cite{PhysRevB.107.075121} (The cuprate data are consistent with the experimental data reported for 2D cuprates~\cite{PhysRevB.85.214527}.).  The results are consistent with the picture that the cuprates are CT insulators, whereas the excitations in the nickelate appear at larger energies, which supports the mixed valence picture~\cite{YShen2022}. We also notice a weak feature at 1 eV for the nickelate and around 2 eV for the cuprates. These correspond to the $dd$ excitations, and these energy scales are consistent with the plaquette picture discussed in appendix~\ref{App:TMO4toOnesite}.   

Panel b) shows the O $K$-edge RIXS spectra for the doped (50\%) nickelate and cuprate. We notice an overall softening of the CT excitations in both. In the case of nickelates, a distinct feature at low energy feature at 1-2 eV appears. This feature has also been interpreted as $dd$ excitations in Ref.~\cite{YShen2022}. For the doped cuprate model, the $dd$ and CT excitations overlap.  \\

\section{Angular dependence}\label{App:RIXSpol}

\begin{figure}
	\centering
	\includegraphics[width=0.25\linewidth]{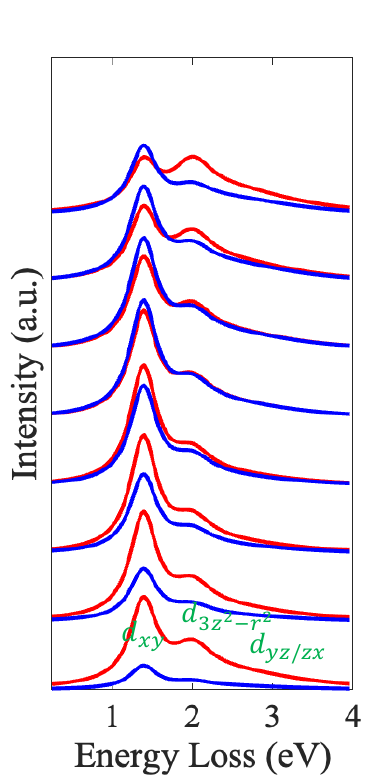}
\caption{Angular dependence for a set of incident energies of dd excitation evaluated using single ion ($d^9$) model for the $\sigma$ (red) and $\pi$ (blue) -polarization. This reproduces the angular dependence of $dd$ excitations for a set of incidence energies reported in Ref~\cite{PhysRevB.104.L220505}}
	\label{fig:angulardependece}
\end{figure}

We revisit the single site model proposed in Ref.~\cite{PhysRevB.104.L220505} for understanding the hierarchy of the $dd$ excitations in nickelates. The $dd$ excitations were identified using the angular dependence of the $d$ orbitals using a single ion picture~\cite{MorettiSala2011}. We notice that the angular dependence for $d_{3z^2}$ and $d_{yz/zx}$ are very close to the set of parameters reported in the paper, allowing for redundancy in the model. Here, we reanalyze the data using an alternate set of parameters; $\epsilon_{x^2-y^2}=0, \epsilon_{xy} (\Gamma_{xy}) =1.39 (=0.2), 
\epsilon_{3z^2}~( \Gamma_{3z^2})=2.0~(0.3), \epsilon_{yz/zx}~(\Gamma_{yz/zx}) = 2.5~(1.0)$ and the approach as used in ref.~\cite{PhysRevB.104.L220505, MorettiSala2011}. We can reproduce the experimentally observed data by interchanging the hierarchy of $3z^2$ and $yz/zx$-orbitals which is shown in Fig.~\ref{fig:angulardependece}. This suggests that the $3z^2$ need not be the farthest from $d_{x^2-y^2}$-orbital and is indeed the lowest energy $dd$ excitation.

\section{Anderson Impurity Model}

The Anderson Impurity model used in our work is given by
\begin{equation}
\begin{split}
H_{AIM}  &= \sum_{\alpha} \epsilon_{\alpha} d_{\alpha}^\dagger d_{\alpha}^\pdag + \frac{1}{2} \sum_{\alpha\beta\gamma\delta} U_{\alpha\beta\gamma\delta} d_{\alpha}^\dagger d_{\beta}^\dagger d_{\gamma}^\pdag d_{\delta}^\pdag \\
& + \sum_{l\alpha} \epsilon_{l\alpha} b_{l\alpha}^\dagger b_{l\alpha}^\pdag + \sum_{l\alpha} V_{l\alpha} d_{\alpha}^\dagger b_{l\alpha}^\pdag +h.c. 
\end{split}
\end{equation}
Here, $\alpha, \beta, \gamma, \delta$ are the orbital indices of the impurity site, which represent the $3d$-orbitals of the Ni. $ l~(=\{1 \cdots N\})$ is the index for the bath site. Due to computational limitations, we restrict to $N=2$ bath sites.
$\epsilon_{l\alpha}$ is the energy level of the $l$-th bath site with orbital $\alpha$, $V_{l\alpha}$ are the hybridization strength between localized impurity electrons $ (d_\alpha^\pdag)$ and bath electrons $(b_{l\alpha}^\pdag)$.
The bath levels and hybridization strengths do not exist explicitly within the DFT+DMFT functional, which integrates out these bath degrees of freedom into a Weiss field with a frequency-dependent hybridization function, $\Delta(i\omega_n)$. Here, $\epsilon_{l\alpha}$ and $V_{l\alpha}$ are obtained by fitting them to the diagonal elements of the hybridization function via the following equation:
\begin{equation}
\Delta_{\alpha\alpha}(i\omega_n) = \sum_{l=1}^{N} \frac{|V_{l\alpha}|^2}{i\omega_n -\epsilon_{l\alpha}}.
\end{equation}
The intermediate Hamiltonian with core is given by $H_n = H_{AIM} +H_\text{ch}$, where the core-holed effects are included in the $3d$ orbitals for the $L$-edge. We evaluate the Ni $L$-edge RIXS spectra for the AIM using EDRIXS. We use exact diagonalization to solve the AIM for both the undoped and doped materials. We restrict ourselves to two bath sites (see Sec.~2.3 in Ref.~\cite{WANG2019151} for details). The calculations are performed at 300 K and use a virtual doping of the lanthanum ion to simulate the doped $(d^{8.75})$ system, and the results are shown in the Fig. 4 of the main text, which contains both the NSC and SC channels.


%
\makeatother

\end{document}